\documentclass[12pt]{article}
\usepackage{graphicx}
\def\bra#1{\left\langle #1\right|}
\def\ket#1{\left| #1\right\rangle}
\def\beq{\begin{equation}}
\def\eeq{\end{equation}}
\def\bea{\begin{eqnarray}}
\def\eea{\end{eqnarray}}
\def\nn{\nonumber}
\def\roughly#1{\mathrel{\raise.3ex\hbox
{$#1$\kern-.75em\lower1ex\hbox{$\sim$}}}}
\def\lsim{\roughly<}

\def\sss{\scriptscriptstyle}

\def\bd{B_d^0}
\def\bdbar{{\overline{B_d^0}}}
\def\bs{B_s^0}
\def\bsbar{{\overline{B_s^0}}}
\def\ks{K_{\sss S}}
\def\kbar{{\bar K}^0}
\def\ssbar{s{\bar s}}

\def\Aut{{\cal A}_{ut}}
\def\Act{{\cal A}_{ct}}
\def\Autp{{\cal A}'_{ut}}
\def\Actp{{{\cal A}'_{ct}}}

\def\ANPdir{{\cal A}_{\sss NP}^{dir}}
\def\ANPr{{\cal A}_{\sss NP}^{rescatt}}
\def\ANPq{{\cal A}_{\sss NP}^q}
\def\ANPu{{\cal A}_{\sss NP}^u}
\def\ANPd{{\cal A}_{\sss NP}^d}
\def\ANPs{{\cal A}_{\sss NP}^s}
\def\ANPc{{\cal A}_{\sss NP}^c}
\def\aI{a_{\sss I}}
\def\aR{a_{\sss R}}

\def\ssbar{(s{\bar s})}
\def\btod{{\bar b} \to {\bar d}}
\def\btos{{\bar b} \to {\bar s}}
\def\npb#1#2#3{{ Nucl.\ Phys.} {\bf B#1}, #3 (#2)}
\def\plb#1#2#3{{ Phys.\ Lett.} {\bf #1B}, #3 (#2)}
\def\prd#1#2#3{{ Phys.\ Rev.} {\bf D#1}, #3 (#2)}

\def\prl#1#2#3{{ Phys.\ Rev.\ Lett.} {\bf #1}, #3 (#2)}

\begin{document}

\begin{flushright}  
UdeM-GPP-TH-04-121 \\
McGill 11/04 \\
\end{flushright}

\begin{center}
\bigskip
{\Large \bf \boldmath Measuring New-Physics Parameters \\ in $B$ Penguin
Decays} \\
\bigskip
\bigskip
{\large Alakabha Datta $^{a,}$\footnote{datta@physics.utoronto.ca} and
David London $^{b,c,}$\footnote{london@lps.umontreal.ca}}
\end{center}

\begin{flushleft}
~~~~~~~~~~~$a$: {\it Department of Physics, University of Toronto,}\\
~~~~~~~~~~~~~~~{\it 60 St.\ George Street, Toronto, ON, Canada M5S 1A7}\\
~~~~~~~~~~~$b$: {\it Physics Department, McGill University,}\\
~~~~~~~~~~~~~~~{\it 3600 University St., Montr\'eal QC, Canada H3A 2T8}\\
~~~~~~~~~~~$c$: {\it Laboratoire Ren\'e J.-A. L\'evesque, 
Universit\'e de Montr\'eal,}\\
~~~~~~~~~~~~~~~{\it C.P. 6128, succ. centre-ville, Montr\'eal, QC,
Canada H3C 3J7}
\end{flushleft}

\begin{center} 
\bigskip (\today)
\vskip0.5cm
{\Large Abstract\\}
\vskip3truemm
\parbox[t]{\textwidth} {We examine new-physics (NP) effects in $B$
decays with large $\btos$ penguin amplitudes. Decays involving $\btod$
penguins are assumed to be unaffected. We consider a model-independent
parametrization of such NP. We argue that NP strong phases are
negligible relative to those of the standard model. This allows us to
describe the NP effects in terms of a small number of effective
amplitudes $\ANPq$ ($q=u,d,s,c$) and corresponding weak phases
$\Phi_q$. We then consider pairs of neutral $B$ decays which are
related by flavour SU(3) in the standard model. One receives a large
$\btos$ penguin component and has a NP contribution; the other has a
$\btod$ penguin amplitude and is unaffected by NP. The time-dependent
measurement of these two decays allows the {\it measurement} of the NP
parameters $\ANPq$ and $\Phi_q$. The knowledge of these parameters
allows us to rule out many NP models and thus partially identify the
new physics.}
\end{center}

\thispagestyle{empty}
\newpage
\setcounter{page}{1}
\baselineskip=14pt

The $B$-factories BaBar and Belle have already made a large number of
measurements involving $B$ decays, and this will continue for a number
of years. The principal aim of this activity is to test whether the
standard model (SM) explanation of CP violation --- a complex phase in
the Cabibbo-Kobayashi-Maskawa (CKM) matrix \cite{pdg} --- is correct.
This is done by measuring CP violation in the $B$ system in many
different processes \cite{CPreview}. Hopefully a discrepancy will be
found, giving us the first indication of physics beyond the SM.

New-physics (NP) effects in $B$ decays are necessarily virtual
processes. As a result, it is generally assumed that, while
$B$-factories can detect the presence of NP, its identification can
only be made at future high-energy colliders, in which the new
particles are produced directly. The main purpose of this paper is to
show that this is not entirely true. Here we will describe a technique
which allows us not only to detect the NP, but also to {\it measure}
its amplitude and phase. This will be an important first step in
identifying the new physics, even before it has been seen directly at
high-energy colliders.

Recently, there have been several hints of such new physics. First,
within the SM, the CP-violating asymmetries in $\bd(t) \to J/\psi \ks$
and $\bd(t) \to \phi\ks$ are both expected to measure the same
quantity $\sin 2\beta$ \cite{phiKsSM}. However, the Belle measurement
of $\sin 2 \beta$ in $\bd(t) \to \phi\ks$ disagrees with that found in
$\bd(t) \to J/\psi \ks$ by $3.5\sigma$ (there is no discrepancy in the
BaBar result) \cite{HFAG}. Indeed, the value of $\sin 2\beta$
extracted from all $\btos$ penguin decays is $3.1\sigma$ below that
from charmonium decays. Second, the various $B\to K\pi$ branching
ratios have been measured. If one neglects exchange- and
annihilation-type amplitudes, which are expected to be small, within
the SM one has $R_c = R_n$ \cite{GroRos}, where
\beq
R_c \equiv \frac{2 {\bar\Gamma}(B^+ \to K^+
\pi^0)}{{\bar\Gamma}(B^+\to K^0\pi^+)} ~~,~~~~
R_n \equiv \frac{{\bar\Gamma}(\bd \to K^+ \pi^-)}{2{\bar\Gamma}(\bd\to
K^0\pi^0)} ~.
\eeq
However, current measurements yield \cite{HFAG}
\beq
R_c = 1.42 \pm 0.18 ~~,~~~~ R_n = 0.89 \pm 0.13 ~,
\eeq
yielding a discrepancy of $2.4\sigma$ between $R_c$ and $R_n$.
Finally, within the SM all CP-violating triple-product correlations
(TP's) in $B\to V_1 V_2$ decays ($V_1$ and $V_2$ are vector mesons)
are expected to vanish or be very small \cite{BVVTP}. However, BaBar
sees a TP signal in $B \to \phi K^*$ at $1.7\sigma$ \cite{TPsignal}.

While the above new-physics signals are not yet convincing, they do
suggest that NP might be playing a role in these decays. In addition,
in all cases, the decays in question ($B \to \phi K^{(*)}$ and $B\to
K\pi$) receive significant contributions from $\btos$ penguin
amplitudes. On the other hand, to date there are no NP signals in
processes which receive sizeable contributions from $\btod$ penguin
amplitudes (e.g.\ $\bd\to \pi\pi$). In this paper, we therefore make
the assumption that NP contributes significantly only to those decays
which have large $\btos$ penguin amplitudes; decays involving $\btod$
penguins are not affected.

Up to now, theoretical work has focussed principally on finding
signals of new physics in $\btos$ transitions -- in fact, there are
many such signals. However, if NP is found, we will want to identify
it. This requires the determination of the NP parameters.
Unfortunately, most NP signals simply indicate that physics beyond the
SM is present, but do not allow us to extract its parameters. (In some
cases, it is possible to put bounds on the NP parameters
\cite{bounds}.) The advantage of the technique described in this paper
is that it allows us to {\it measure} the amplitude and phase of the
NP.

Assuming that the new physics affects only the $\btos$ penguin
amplitudes, the first step is a model-independent parametrization of
this NP. We assume that a NP piece is added to the effective
Hamiltonian:
\beq
H_{eff} = H_{\sss SM}+H_{\sss NP} ~,
\eeq
where $H_{\sss SM}$ is the SM effective Hamiltonian \cite{BuraseffH}.
$H_{\sss NP}$ contains four-quark operators with all possible Dirac
and colour structures, with the proviso that only $\btos$ penguin
transitions are affected. That is, the general structure of the
operators in $H_{\sss NP}$ is $O_{\sss NP} \sim \bar{s} b \, \bar{q}
q$ ($q=u,d,s,c$), where Lorentz and colour structures have been
suppressed. We also assume that the contribution from $O_{\sss NP}$ to
any $B$ decay is at most of the same order as the SM penguin
amplitude.

Taking into account the two different colour assignments, as well as
all possible Lorentz structures, there are a total of 20 dimension-six
new-physics operators which contribute to each of the $\btos q\bar{q}$
($q=u,d,s,c$) transitions \cite{lambdabNP}. These operators, which can
contribute to both tree and penguin amplitudes, can be written as
\bea
{\cal H}^q_{\sss NP} = \sum_{\sss A,B = L,R} {4 G_{\sss F} \over
\sqrt{2}} &&
\hskip-5truemm
\left\{ f_{q,1}^{\sss AB} \, {\bar s}_\alpha \gamma_{\sss A} b_\beta
\, {\bar q}_{\beta} \gamma_{\sss B} q_{\alpha} + f_{q,2}^{\sss AB} \,
{\bar s} \gamma_{\sss A} b \, {\bar q} \gamma_{\sss B} q \right. \nn\\
&& \hskip-7truemm 
+ ~ g_{q,1}^{\sss AB} \, {\bar s}_\alpha \gamma^\mu \gamma_{\sss A}
b_\beta \, {\bar q}_{\beta} \gamma_\mu \gamma_{\sss B} q_{\alpha} +
g_{q,2}^{\sss AB} \, {\bar s} \gamma^\mu \gamma_{\sss A} b \, {\bar q}
\gamma_\mu \gamma_{\sss B} q \nn\\
&& \hskip-7truemm \left.
+ ~ h_{q,1}^{\sss AB} \, {\bar s}_\alpha \sigma^{\mu\nu} \gamma_{\sss
A} b_\beta \, {\bar q}_{\beta} \sigma_{\mu\nu} \gamma_{\sss B}
q_{\alpha} + h_{q,2}^{\sss AB} \, {\bar s} \sigma^{\mu\nu}
\gamma_{\sss A} b \, {\bar q} \sigma_{\mu\nu} \gamma_{\sss B} q
\right\},
\label{HeffNP}
\eea
where we have defined $\gamma_{\sss R(L)}= {1\over 2} (1 \pm
\gamma_5)$. Although we have written the tensor operators in the same
compact form as the other operators, it should be noted that those
with $\gamma_{\sss A} \ne \gamma_{\sss B}$ are identically zero. Thus,
one can effectively set $h_{q,i}^{\sss LR} = h_{q,i}^{\sss RL} = 0$.

In general, all coefficients in Eq.~(\ref{HeffNP}) can have new
CP-violating weak phases and the matrix elements of the operators will
have (process-dependent) CP-conserving strong phases. Given the large
number of possible operators it is virtually impossible to isolate the
amplitudes and phases of the different operators. (It may be possible
to do this in the context of a particular model, in which only a small
subset of operators is present.) Fortunately, as we argue below, the
strong phases of all NP operators are small relative to those of the
SM and can be neglected. As a result, the various NP terms can be
combined into a single NP operator, whose amplitude and phase {\it
can} be measured.

To see how this works, consider $\bd\to\phi\ks$. (This is chosen for
illustration only -- the argument holds for any $B$ decay which
receives a significant $\btos$ penguin contribution in the SM, and is
dominated by a single amplitude.) The SM amplitude for this decay can
be written
\beq
A(\bd\to\phi\ks) = A'_u V_{ub}^* V_{us} + A'_c V_{cb}^* V_{cs} + A'_t
V_{tb}^* V_{ts} ~.
\label{BphiKamp}
\eeq
Here, $A'_t$ arises due to the gluonic penguin amplitude with a
$t$-quark in the loop. Although $A'_u$ and $A'_c$ also receive (small)
contributions from the gluonic penguin, they arise mainly as a result
of QCD rescattering from the tree operators $\btos u{\bar u}$ and
$\btos c {\bar c}$. The Wilson coefficients for the various
contributions imply that $A'_u,A'_c \lsim 0.5 A'_t$. Note that the
size of the rescattered penguin amplitudes is only about 5--10\% of
that of the tree amplitude. Using CKM unitarity, the amplitude for
$\bd\to\phi\ks$ can be written
\beq
A(\bd\to\phi\ks) = \Autp\ e^{i \gamma} e^{i \delta'_{ut}} + \Actp\ e^{i
\delta'_{ct}} \approx \Actp\ e^{i \delta'_{ct}} ~,
\eeq
where $\Autp \equiv |(A'_u - A'_t) V_{ub}^* V_{us}|$ and $\Actp \equiv
|(A'_c - A'_t) V_{cb}^* V_{cs}|$. The final (approximate) equality
arises from the fact that $\left\vert {V_{ub}^* V_{us} / V_{cb}^*
V_{cs}} \right\vert \simeq 2\%$, so that $\Autp \ll \Actp$. The
quantity $\delta'_{ct}$ is a strong phase; the weak phase is
approximately zero.

The principal NP contribution to $\bd\to\phi\ks$ comes from $\bar{s}b
\bar{s} s$ (both Lorentz and colour factors are once again
suppressed). However, other NP operators, such as $\bar{s}b \,
\bar{c} c$, can also contribute to $\bd\to\phi\ks$ through
rescattering. The full amplitude for this decay can therefore be
written
\bea
& A_{\phi \ks} = \Actp\ e^{i \delta'_{ct}} 
+ \ANPdir +\ANPr ~, & \nn\\
& \ANPdir \equiv \sum_i A_i e^{i \phi_i^{ss}} e^{i \delta_i} ~~,~~~~ 
\ANPr \equiv \sum_i \epsilon_i B_i e^{i \xi_i} e^{i \sigma_i} ~. &
\label{generalNPamp}
\eea
In the above, $\ANPdir$ is the contribution from all NP operators of
the form $\bar{s} \Gamma_i b \, \bar{s} \Gamma_j s$ ($\Gamma_{i,j}$
represent Lorentz structures, and colour indices are suppressed),
while $\ANPr$ is the contribution from all NP operators of the form
$\bar{s} \Gamma_i b \, \bar{q} \Gamma_j q$ ($q \ne s$). In the latter
case, the decays $\btos q {\bar q}$ ($q \ne s$) contribute to $\btos s
{\bar s}$ through rescattering. Similarly, $\ANPdir$ includes the
``self-rescattering'' contributions of $\btos s {\bar s}$ to $\btos s
{\bar s}$. The NP weak phases are $\phi_i^s$ and $ \xi_i$, while
$\delta_i$ and $\sigma_i$ are the NP strong phases.

At this point, it is useful to discuss rescattering in somewhat more
detail. As noted above, in the SM, for decays described by $\btos$
transitions, the rescattering comes mainly from the tree-level decay
$\btos c {\bar c}$. Although the rescattered ``penguin'' amplitudes
$A'_u$ and $A'_c$ are only about 5--10\% as large as the amplitude
which causes the rescattering, they are still of the same order as
$A'_t$ [see Eq.~(\ref{BphiKamp})]. That is, the SM rescattering
effects are not small. In particular, since it is rescattering which
is the principal source of strong phases, the phase $\delta'_{ct}$ in
Eq.~(\ref{generalNPamp}) can be sizeable.

Now, the new-physics rescattering arises from the NP operators. As in
the SM, the rescattered amplitude is suppressed by $\epsilon_i \sim $
5--10\% relative to the operator causing the recattering. Thus,
although $B_i \sim A_i$ in Eq.~(\ref{generalNPamp}), $|\ANPr|$ is only
5--10\% as large as $|\ANPdir|$. (The rescattered contributions in
$\ANPdir$ are similarly suppressed.) However, the NP operators are
assumed to be of the same size as the SM $\btos$ penguin amplitude
$\Actp$ [Eq.~(\ref{generalNPamp})]. Therefore $\ANPr$ is negligible
compared to $\Actp$ and $\ANPdir$. In addition, we note that the NP
strong phase $\delta_i$ in $\ANPdir$ vanishes in the limit of no
rescattering. Since, as we have argued, this NP rescattering is small,
we have $\delta_i \ll \delta'_{ct}$, i.e.\ the NP strong phases are
negligible compared to those of the SM.

These approximations lead to a considerably simpler structure for
Eq.~(\ref{generalNPamp}):
\bea
A_{\phi \ks} & \approx & \Actp\ e^{i \delta'_{ct}} + \ANPdir ~, \nn\\
\ANPdir & \equiv & \sum_i A_i e^{i \phi_i^s}= \ANPs e^{ i \Phi_s} ~,
\label{generalNPampdom}
\eea
where we have summed up the new physics contributions into a single
amplitude. The important point here is that all the NP weak phases
come only from operators of the type $O_{\bar{s} s}= \bar{s} \Gamma_i
b \, \bar{s} \Gamma_j s$, and so the effective weak phase carries the
subscript ``$s$'': $\Phi_s$. {}From Eq.~(\ref{generalNPampdom}) we
have
\bea
\tan{ \Phi_s} & = & \frac{ \sum_i A_i \sin{\phi_i^s}} 
{ \sum_i A_i \cos{\phi_i^s}} ~.
\label{phase_ss}
\eea

The above argument holds for the case where there are new-physics
contributions to $\btos q {\bar q}$ ($q=d,s,c$). However, $\btos u
{\bar u}$ is slightly different because the SM decay is not dominated
by a single amplitude -- there are both tree and penguin
contributions. Nevertheless, it is straightforward to show that the
above logic still holds: the rescattering in the NP amplitudes to
$\btos u {\bar u}$ is negligible, so that the NP contributions can be
parametrized by a single amplitude $\ANPu$ and weak phase $\Phi_u$.

Thus, under the assumption that new-physics rescattering is negligible
compared to that of the SM, the effects of the NP operators $\bar{s}b
\bar{q} q$ can be parametrized in terms of the effective NP amplitudes
$\ANPq$ ($q=u,d,s,c$) and the corresponding weak phases $\Phi_q$. In
the rest of the paper we will show how these NP parameters can be
measured.

Note that there may be a possible loophole in the above argument. In
the SM, the exchange and annihilation contributions are expected to be
quite small, for both $\btod$ and $\btos$ transitions. However, in
some approaches to hadronic $B$ decays, such amplitudes may be
chirally enhanced if there are pseudoscalars in the final state
\cite{BBNS,PQCD}, with resulting large strong phases. Hence
annihilation-type topologies generated by NP operators may also lead
to large strong phases. On the other hand, such chiral enhancements
are not present for vector-vector final states and so the above
arguments regarding small NP strong phases are applicable here.
Ultimately, the size of exchange and annihilation diagrams is an
experimental question, and can be tested by the measurement of decays
such as $\bd \to D_s^+D_s^-$ and $\bd \to K^+ K^-$.

In general, we take the effective new-physics phases $\Phi_q$ to be
flavour non-universal. That is, we assume that the phases for
different underlying quark transitions, $\btos q {\bar q}$, are not
related. This occurs in many models of NP, such as supersymmetry with
R-parity-violating terms \cite{phiKsNPRpar}. However, there are also
NP models, such as those including a flavour-changing $Z$ or $Z'$
coupling \cite{phiKsNPZFCNC}, in which the phases {\it are}
related. This shows that the measurement of the $\Phi_q$ will be
very useful in identifying the new physics, or at least excluding
certain NP models. Note that if {\it all} the NP operators have the same
weak phase $\varphi$, one has $\Phi_q=\varphi$, and this phase
is process-universal as well as flavour-universal. In this case one
can simplify Eq.~\ref{generalNPamp} as
\bea
& A_{\phi \ks} = \Actp\ e^{i \delta'_{ct}} +\ANPs\ e^{ i \delta_{\sss
NP}}\ e^{ i \varphi}, & \nn\\
& \ANPs\ e^{ i \delta_{\sss NP}} = \sum_i A_i e^{i \delta_i}+ \sum_i
\epsilon_iB_i e^{i \sigma_i} ~. &
\label{generalNPamp_onephase}
\eea
Factoring out the strong phase $\delta_{\sss NP}$, it is clear that we
can cast Eq.~(\ref{generalNPamp_onephase}) in the same form as
Eq.~(\ref{generalNPampdom}) without any dynamical input about NP
strong phases. 

Above we showed that the new-physics effects can be parametrized in
terms of a few effective NP parameters. We now describe a method for
{\it measuring} these parameters. This technique closely resembles
that of Ref.~\cite{Bpenguin}, which we recently proposed for
extracting CP phase information. Here we turn this method around. As
above, we assume that NP is present only in decays with large $\btos$
penguin amplitudes. We further assume that the SM CP phase information
is known: these phases can be measured using processes which do not
involve large $\btos$ penguin amplitudes. In this case, the method can
be used to extract the NP parameters.

In order to illustrate the method, we consider a specific pair of $B$
decays. It is straightforward to adapt the technique to other
processes. Consider $\bs \to K^0 \kbar$. In the SM, this decay is
dominated by a single $\btos d{\bar d}$ penguin decay amplitude.
Including new physics, the amplitude for $\bs \to K^0 \kbar$ can be
written as [see Eq.~(\ref{generalNPampdom})]
\beq
A(\bs \to K^0 \kbar) \equiv A = \Actp\ e^{i \delta'_{ct}} + \ANPd e^{i
\Phi_d} ~,
\label{generalNPampdom1}
\eeq
where $\Actp$ and $\ANPd$ are the SM and NP amplitudes, respectively.
Similarly, $\delta'_{ct}$ and $\Phi_d$ are the SM strong phase and NP
weak phase, respectively. The NP phase is defined analogously to
Eq.~(\ref{phase_ss}). The amplitude for the CP-conjugate process,
${\overline A}$, can be obtained from the above by changing the sign
of $\Phi_d$.

Since the final state $K^0\kbar$ is accessible to both $\bs$ and
$\bsbar$ mesons, one can consider indirect (mixing-induced) CP
violation. The time-dependent measurement of $\bs(t) \to K^0\kbar$
allows one to obtain the three observables
\bea
B &\equiv & \frac{1}{2} \left( |A|^2 + |{\overline A}|^2 \right) =
(\Actp)^2 + (\ANPd)^2 + 2 \Actp \, \ANPd \cos\delta'_{ct}
\cos\Phi_d ~, \nn \\
\label{meas1}
a_{dir} &\equiv & \frac{1}{2} \left( |A|^2 - |{\overline A}|^2 \right)
= 2 \Actp \, \ANPd \sin\delta'_{ct} \sin\Phi_d ~, \\
\aI &\equiv & {\rm Im}\left( e^{-2i \phi_{\sss B_s}} A^* {\overline A}
\right) = -(\Actp)^2 \sin 2\phi_{\sss B_s} - 2 \Actp \, \ANPd
\cos\delta'_{ct} \sin (2 \phi_{\sss B_s} + \Phi_d) \nn\\
& & \hskip2.6truein
- (\ANPd)^2 \sin (2\phi_{\sss B_s} + 2 \Phi_d)~. \nn
\eea
It is useful to define a fourth observable:
\bea
\aR & \equiv & {\rm Re}\left( e^{-2i \phi_{\sss B_s}} A^* {\overline A}
\right) = (\Actp)^2 \cos 2\phi_{\sss B_s} + 2 \Actp \, \ANPd
\cos\delta'_{ct} \cos (2 \phi_{\sss B_s} + \Phi_d) \nn\\
& & \hskip2.6truein
+ (\ANPd)^2 \cos (2\phi_{\sss B_s} + 2 \Phi_d)~.
\label{meas2}
\eea
The quantity $\aR$ is not independent of the other three observables:
\beq
{\aR}^2 = {B}^2 - {a_{dir}}^2 - {\aI}^2 ~.
\label{aRdefNP}
\eeq
Thus, one can obtain $\aR$ from measurements of $B$, $a_{dir}$ and
$\aI$, up to a sign ambiguity.

In the above, $\phi_{\sss B_s}$ is the phase of $\bs$--$\bsbar$
mixing. In general, NP which affects $\btos$ transitions will also
contribute to $\bs$--$\bsbar$ mixing, i.e.\ one will have NP operators
of the form $\bar{s} b \, \bar{b} s$. In this case, the phase of
$\bs$--$\bsbar$ mixing may well differ from its SM value ($\simeq 0$)
due to the presence of NP. The standard way to measure this mixing
phase is through CP violation in $\bs(t)\to J/\psi \eta$ (or
$\bs(t)\to J/\psi \phi$). However, there is a potential problem here:
this decay receives NP contributions from $O_{\sss NP}^c \sim \bar{s}b
{\bar c} c$ operators (as usual, the Lorentz and colour structures
have been suppressed), so that there may be effects from these NP
operators in any process involving $\bs$--$\bsbar$ mixing.

The solution to this problem can be found by considering
$\bd$--$\bdbar$ mixing. The phase of this mixing is unaffected by new
physics and thus takes its SM value, $\beta$. The canonical way to
measure this angle is via CP violation in $\bd(t)\to J/\psi \ks$.
However, this decay also receives NP contributions from $O_{\sss
NP}^c$ operators. On the other hand, the value of $\beta$ extracted
from $\bd(t)\to J/\psi \ks$ is in line with SM expectations. This
strongly suggests that any $O_{\sss NP}^c$ contributions to this decay
are quite small. Now, the non-strange part of the $\eta$ wavefunction
has a negligible contribution to $\bra{J/\psi \eta}O_{\sss NP}^c
\ket{\bs}$. Thus, this matrix element can be related by flavour SU(3)
to $\bra{J/\psi \ks}O_{\sss NP}^c \ket{\bd}$ (up to a mixing
angle). That is, both matrix elements are very small. In other words,
we do not expect significant $O_{\sss NP}^c$ contributions to
$\bs(t)\to J/\psi \eta$, and the phase of $\bs$--$\bsbar$ mixing can
be measured through CP violation in this decay, even in the presence
of NP.

We have already noted that there are many signals of new physics in
$B$ decays. Indeed, the expressions for $a_{dir}$ and $\aI$ in
Eq.~(\ref{meas1}) give us clear signals of NP. Since $\bs\to K^0\kbar$
is dominated by a single decay in the SM, the direct CP asymmetry is
predicted to vanish. Furthermore, the indirect CP asymmetry is
expected to measure the mixing phase $\phi_{\sss B_s} \simeq 0$. Thus,
if it is found that $a_{dir} \ne 0$, or that $\phi_{\sss B_s}$ does
not take its SM value, this would be a smoking-gun signal of NP. Note
also that, if it happens that the SM strong phases are small,
$a_{dir}$ may be unmeasurable. In this case, a better signal of new
physics is the measurement of T-violating triple-product correlations
in the corresponding vector-vector final states \cite{BVVTP}. This
brief discussion illustrates that there are indeed many ways of
detecting the presence of NP. However, it must also be stressed that
these signals do not, by themselves, allow the measurement of the NP
parameters.

The three independent observables of Eqs.~(\ref{meas1}) and
(\ref{meas2}) depend on four unknown theoretical parameters: $\ANPd$,
$\Actp$, $\delta'_{ct}$ and $\Phi_d$. Therefore one cannot obtain
information about the new-physics parameters $\ANPd$ and $\Phi_d$ from
these measurements. However, one can partially solve the equations to
obtain
\beq
(\Actp)^2 = { \aR \cos(2\phi_{\sss B_s} + 2\Phi_d) - \aI
\sin(2\phi_{\sss B_s} + 2\Phi_d) - B \over \cos 2\Phi_d - 1} ~.
\label{phicond}
\eeq
Thus, if we knew $\Actp$, we could solve for $\Phi_d$.

In order to get $\Actp$ we consider the partner process $\bd\to
K^0\kbar$, involving a $\btod$ penguin amplitude. In the SM this decay
is related by SU(3) symmetry to $\bs\to K^0\kbar$ \cite{alphapaper}.
Since $\btos$ transitions are not involved, the amplitude for $\bd\to
K^0\kbar$ receives only SM contributions, and is given by
\bea
A(\bd\to K^0\kbar) & = & A_u V_{ub}^* V_{ud} + A_c V_{cb}^*
V_{cd} + A_t V_{tb}^* V_{td} \nn\\
& = & (A_u - A_t) V_{ub}^* V_{ud} + (A_c - A_t) V_{cb}^* V_{cd} \nn\\
& \equiv & \Aut\ e^{i \gamma} e^{i \delta_{ut}} + \Act\ e^{i
  \delta_{ct}} ~,
\label{Bfamp}
\eea
where $\Aut \equiv |(A_u - A_t) V_{ub}^* V_{ud}|$, $\Act \equiv |(A_c
- A_t) V_{cb}^* V_{cd}|$, and we have explicitly written the strong
phases $\delta_{ut}$ and $\delta_{ct}$, as well as the weak phase
$\gamma$. 

As with $\bs\to K^0 \kbar$, the time-dependent measurement of
$\bd(t)\to K^0 \kbar$ allows one to obtain three independent
observables [Eqs.~(\ref{meas1}) and (\ref{meas2})]. These observables
depend on five theoretical quantities: $\Act$, $\Aut$, ${\delta}\equiv
{\delta}_{ut} - {\delta}_{ct}$, $\gamma$ and the mixing phase
$\phi_{\sss B_d}$. However, as discussed above, $\phi_{\sss B_d}$ can be
measured independently using $\bd(t)\to J/\psi \ks$. The weak phase
$\gamma$ can also be measured in $B$ decays which are unaffected by
new physics in $\btos$ penguin amplitudes. For example, it can be
obtained from $B^\pm \to DK$ decays \cite{BDK}. Alternatively, the
angle $\alpha$ can be extracted from $B \to \pi\pi$ \cite{Bpipi} or
$B\to\rho\pi$ decays \cite{Brhopi}, and $\gamma$ can be obtained using
$\gamma = \pi - \beta - \gamma$. Given that these CP phases can be
measured independently, the three observables of $\bd(t)\to K^0\kbar$
now depend on three unknown theoretical parameters, so that the system
of equations can be solved.

In particular, one can obtain $\Act$:
\beq
\Act^2 = { \aR \cos(2\phi_{\sss B_d} + 2\gamma) - \aI \sin(2\phi_{\sss
\bd} + 2\gamma) - B \over \cos 2\gamma - 1} ~,
\label{gammacondrevisited}
\eeq
where $\aR$, $\aI$ and $B$ are the observables found in $\bd(t) \to
K^0\kbar$.

The key point is that, in the SU(3) limit, one has
\beq
\Act=\lambda \Actp ~,
\label{Actrel}
\eeq
where $\lambda = 0.22$ is the Cabibbo angle. Thus, using the above
relation, the measurement of $\bd(t)\to K^0 \kbar$ gives us $\Actp$,
in which case Eq.~(\ref{phicond}) can be used to solve for the new
physics phase $\Phi_d$. The NP amplitude $\ANPd$ can also be
obtained. There is a theoretical error in Eq.~(\ref{Actrel}) due to
SU(3)-breaking effects. However, various methods were discussed in
Ref.~\cite{Bpenguin} to reduce this SU(3) breaking. All of these
methods are applicable here. In the end, for this particular pair of
processes, the theoretical error is estimated to be in the range
5--10\%.

Above, we have shown how measurements of the decays $B_{d,s}^0(t) \to
K^0\kbar$ can be used to measure the NP parameters $\ANPd$ and
$\Phi_d$. The general idea is to use a $\btos$ decay which is
dominated in the SM by a single decay amplitude, along with its
$\btod$ partner process. This method can be adapted to other pairs of
$B$ decays to measure different NP parameters. (Or one can find
alternative ways of measuring $\ANPd$ and $\Phi_d$.) By choosing the
two decays carefully, the theoretical error can be reduced to the
level of 5--15\%.

Note that it is only quark-level decays $\btos q{\bar q}$ ($q=d,s,c$)
which are dominated by a single decay amplitude in the SM. However,
one can also apply this technique to $\btos u{\bar u}$ decays, for
which the $\btos$ decay receives both tree and penguin contributions
in the SM. For example, one can use the pair of decays $\bs(t) \to K^+
K^-$ and $\bd(t) \to \pi^+ \pi^-$ to extract the NP parameters $\ANPu$
and $\Phi_u$. However, in this case the theoretical error is
considerably larger since one has to make three SU(3) assumptions of
the type in Eq.~(\ref{Actrel}).

In Table~\ref{summarytable}, we present the list of all $B$ decay
pairs to which this method can be applied, along with the NP
parameters measured. From this Table, we see that all NP parameters
can be obtained. A more detailed analysis of these decays is presented
in Ref.~\cite{DILPSS}. Note that only one decay pair in
Table~\ref{summarytable} involves only $\bd$ decays. The others will
require the time-dependent measurement of $\bs$ decays. However, this
may be difficult experimentally, as $\bs$--$\bsbar$ mixing is large.
For this reason the decay pair $\bd(t) \to K^{*0}\rho^0$ and $\bd(t)
\to\rho^0\rho^0$ may be the most promising for measuring NP
parameters.

\begin{table}
\hfil
\vbox{\offinterlineskip
\halign{&\vrule#&
 \strut\quad#\hfil\quad\cr
\noalign{\hrule}
height2pt&\omit&&\omit&&\omit&\cr
& NP Parameters && $\btos$ Decay && $\btod$ Decay & \cr
height2pt&\omit&&\omit&&\omit&\cr
\noalign{\hrule}
height2pt&\omit&&\omit&&\omit&\cr
& $\Phi_{cc}$, $\ANPc$ && $\bs(t) \to D_s^+ D_s^-$ && $\bd(t) \to D^+
D^-$ & \cr
height2pt&\omit&&\omit&&\omit&\cr
\noalign{\hrule}
height2pt&\omit&&\omit&&\omit&\cr
& $\Phi_s$, $\ANPs$ && $\bd(t) \to\phi K^{*0}$ && $\bs(t) \to \phi
{\bar K}^{*0}$ & \cr
& \omit && $\bs(t) \to \phi \phi$ && $\bs(t) \to \phi {\bar K}^{*0}$ &
\cr
height2pt&\omit&&\omit&&\omit&\cr
\noalign{\hrule}
height2pt&\omit&&\omit&&\omit&\cr
& $\Phi_d$, $\ANPd$ && $\bs(t) \to K^0 {\bar K}^0$ && $\bd(t) \to
\pi^+ \pi^-$ & \cr
& \omit && $\bs(t) \to K^0 {\bar K}^0$ && $\bd(t) \to K^0 {\bar K}^0$
& \cr
& \omit && $\bd(t) \to K^{*0}\rho^0$ && $\bd(t) \to\rho^0\rho^0$ & \cr
& \omit && $\bd(t) \to K^{*0}\rho^0$ && $\bs(t) \to {\bar K}^{*0}
\rho^0$ & \cr
height2pt&\omit&&\omit&&\omit&\cr
\noalign{\hrule}
height2pt&\omit&&\omit&&\omit&\cr
& $\Phi_u$, $\ANPu$ && $\bs(t) \to K^+ K^-$ && $\bd(t) \to \pi^+
\pi^-$ & \cr
height2pt&\omit&&\omit&&\omit&\cr
\noalign{\hrule}}}
\caption{The $\btos$ $B$ decays and their $\btod$ partner processes
which can be used to measure the new-physics parameters $\ANPq$ and
$\Phi_q$.}
\label{summarytable}
\end{table}

Table~\ref{summarytable} lists 8 pairs of $B$ decays. In fact, there
are more decay pairs, since many of the particles in the final states
can be observed as either pseudoscalar (P) or vector (V) mesons. Note
that certain decays are written in terms of VV final states, while
others are have PP states. There are three reasons for this. First,
some decays involve a final-state $\pi^0$. However, experimentally it
will be necessary to find the decay vertices of the final
particles. This is virtually impossible for a $\pi^0$, and so we
always use a $\rho^0$. Second, some pairs of decays are related by
SU(3) in the SM only if an $\ssbar$ quark pair is used. However, there
are no P's which are pure $\ssbar$. The mesons $\eta$ and $\eta'$ have
an $\ssbar$ component, but they also have significant $(u \bar u)$ and
$(d \bar d)$ pieces. As a result the $\btos$ and $\btod$ decays are
not really related by SU(3) in the SM if the final state involves an
$\eta$ or $\eta'$. We therefore consider instead the vector meson
$\phi$ which is essentially a pure $\ssbar$ quark state. Finally, we
require that both $B^0$ and ${\bar B}^0$ be able to decay to the final
state. This cannot happen if the final state contains a single $K^0$
(or ${\bar K}^0$) meson. However, it can occur if this final-state
particle is an excited neutral kaon. In this case one decay involves
$K^{*0}$, while the other has ${\bar K}^{*0}$. Assuming that the
vector meson is detected via its decay to $\ks \pi^0$ (as in the
measurement of $\sin 2\beta$ via $\bd(t) \to J/\psi K^*$), then both
$B^0$ and ${\bar B}^0$ can decay to the same final state.

Apart from these three restrictions, the final-state particles can be
taken to be either pseudoscalar or vector. Indeed, it will be useful
to measure the NP parameters in modes with PP, PV and VV final-state
particles, since different NP operators are probed in these decays.
For example, within factorization, certain scalar operators in
Eq.~\ref{HeffNP} (i.e.\ those whose coefficients are
$f_{q,(1,2)}^{\sss AB}$) cannot contribute to PV or VV states if their
amplitudes involve the matrix element $\bra{V} \bar{q} \gamma_{L,R} q
\ket{0}$. In general, the matrix element of a given operator will be
different for the various PP, PV and VV final states. Thus, the
measurement of the NP parameters in different modes will provide some
clues as to which NP operators are present.

Note also that, in general, the value of $\Phi_q$ extracted from two
distinct decay pairs with the same underlying $\btos q \bar{q}$
transition will be different. There are two reasons for this. First,
certain operators which contribute to one process may not contribute
in the same form in another. (For example, one decay might be
colour-suppressed, while the other is colour-allowed.) Second, in
general, the matrix elements $A_i$ of the various operators depend on
the final states considered.  Thus, the value of the NP phase $\Phi_q$
depends on the particular decay pair used. However, if all NP
operators for the quark-level process $\btos q \bar{q}$ have the same
weak phase $\phi^q$, then the NP phase $\Phi_q$ will be the same for
all decays governed by the same quark-level process. Hence it is
important to measure the phase $\Phi_q$ in more than one pair of
processes with the same underlying quark transition. If the effective
phases are different then it would be a clear signal of more than one
NP amplitude, with different weak phases, in $\btos q \bar{q}$.

It is also important to measure the NP phases $\Phi_q$ for each of
$q=u,d,s,c$. As noted earlier, in some NP models, the phases for the
different underlying quark transitions $\btos q \bar{q}$ are related,
so that the NP phase is independent of the quark flavour. The
measurement of the $\Phi_q$ would thus allow us to distinguish between
NP models.

In summary, it is well known that there are many signals of new
physics (NP) which can be found by measuring CP violation in the $B$
system. However, it is usually assumed that one cannot identify the NP
-- this will have to wait for high-energy colliders which can produce
the new particles directly. In this paper we have shown that this is
not completely true. We have presented a technique which allows the
{\it measurement} of NP parameters. 

In line with hints from present data, we assume that the new physics
contributes only to decays with large $\btos$ penguin amplitudes,
while decays involving $\btod$ penguins are not affected. The NP
rescattering effects are shown to be small compared to those of the SM
and are neglected. This allows us to greatly simplify the form of the
NP contributions. In particular, independent of the type of underlying
NP, we can parametrize all NP effects in terms of effective NP
amplitudes $\ANPq$ and weak phases $\Phi_q$ ($q=u,d,s,c$).

We have shown that one can obtain each of the $\ANPq$ and $\Phi_q$ by
using measurements of pairs of $B$ decays. One decay has a large
$\btos$ penguin component and is (usually) dominated by a single
amplitude. It receives a new-physics contribution. The partner process
has a $\btod$ penguin contribution and is related to the first decay
by flavour SU(3) in the SM. It is unaffected by NP. Assuming that the
SM CP phases are known independently, the measurements of these two
$B$ decays allow one to extract $\ANPq$ and $\Phi_q$. The theoretical
error due to SU(3) breaking can be reduced to the level of 5--15\% for
$q=d,s,c$, but is larger for $q=u$.

In general, different NP models lead to different patterns of the NP
parameters $\ANPq$ and $\Phi_q$. Thus, the measurement of the NP
parameters can rule out certain models and point towards others. We
will therefore have a partial identification of the NP, before
measurements at high-energy colliders.

\bigskip
\noindent
{\bf Acknowledgements}:
We thank M. Gronau and J.L. Rosner for helpful conversations. This
work was financially supported by NSERC of Canada.



\begin{thebibliography}{99}

\bibitem{pdg} K.~Hagiwara {\it et al.}  [Particle Data Group
  Collaboration], Phys.\ Rev.\ D {\bf 66}, 010001 (2002),
  http://pdg.lbl.gov/pdg.html.

\bibitem{CPreview} For a review, see, for example, {\it The BaBar
  physics book: Physics at an asymmetric B factory}, eds.\
  P.~F.~Harrison and H.~R.~Quinn [BABAR Collaboration], SLAC-R-0504,
  October 1998. {\it Papers from Workshop on Physics at an Asymmetric
  B Factory (BaBar Collaboration Meeting), Rome, Italy, 11-14 Nov
  1996, Princeton, NJ, 17-20 Mar 1997, Orsay, France, 16-19 Jun 1997
  and Pasadena, CA, 22-24 Sep 1997.}

\bibitem{phiKsSM} D.~London and R.~D.~Peccei, Phys.\ Lett.\ B {\bf
  223}, 257 (1989); Y.~Nir and H.~R.~Quinn, Ann.\ Rev.\ Nucl.\ Part.\
  Sci.\ {\bf 42}, 211 (1992); Y.~Grossman and M.~P.~Worah, Phys.\
  Lett.\ B {\bf 395}, 241 (1997); D.~London and A.~Soni, Phys.\ Lett.\
  B {\bf 407}, 61 (1997); Y.~Grossman, G.~Isidori and M.~P.~Worah,
  Phys.\ Rev.\ D {\bf 58}, 057504 (1998).

\bibitem{HFAG} The experimental data is tabulated by the Heavy Flavor
  Averaging Group, http://www.slac.stanford.edu/xorg/hfag/.

\bibitem{GroRos} See M.~Gronau and J.~L.~Rosner, Phys.\ Lett.\ B {\bf
  572}, 43 (2003), and references therein.

\bibitem{BVVTP} A.~Datta and D.~London, arXiv:hep-ph/0303159.

\bibitem{TPsignal} Talk by Jim Smith at Moriond, 2004: \\
  http://moriond.in2p3.fr/QCD/2004/TuesdayMorning/Smith.pdf.

\bibitem{bounds} D.~London, N.~Sinha and R.~Sinha,
  arXiv:hep-ph/0304230, \\ arXiv:hep-ph/0402214.

\bibitem{BuraseffH} See, for example, G. Buchalla, A.J. Buras and
  M.E. Lautenbacher, {\it Rev.\ Mod.\ Phys.} {\bf 68}, 1125 (1996),
  A.J. Buras, ``Weak Hamiltonian, CP Violation and Rare Decays,'' in
  {\it Probing the Standard Model of Particle Interactions}, ed.\
  F. David and R. Gupta (Elsevier Science B.V., 1998), pp.\ 281-539.

\bibitem{lambdabNP} W. Bensalem, A. Datta and D. London,
  \prd{66}{2002}{094004}.

\bibitem{BBNS} M. Beneke, G. Buchalla, M. Neubert and C.T. Sachrajda,
  \prl{83}{1999}{1914}, \npb{591}{2000}{313}, \npb{606}{2001}{245}.

\bibitem{PQCD} Y.Y.~Keum, H.-n.~Li and A.I.~Sanda, \plb{504}{2001}{6}.

\bibitem{phiKsNPRpar} The contributions of SUSY models with R-parity
  breaking to $\btos$ transitions, usually in the context of $\bd(t)
  \to \phi \ks$), are examined in A.~Datta, Phys.\ Rev.\ D {\bf 66},
  071702 (2002); B.~Dutta, C.~S.~Kim and S.~Oh, Phys.\ Rev.\ Lett.\
  {\bf 90}, 011801 (2003); A.~Kundu and T.~Mitra, Phys.\ Rev.\ D {\bf
  67}, 116005 (2003);

\bibitem{phiKsNPZFCNC} The model with $Z$-mediated flavour-changing
  neutral currents (FCNC's) was first introduced in Y.~Nir and
  D.~J.~Silverman, Phys.\ Rev.\ D {\bf 42}, 1477 (1990). The
  contributions of $Z$-mediated FCNC's to $\btos$ transitions, usually
  in the context of $\bd(t) \to \phi \ks$), are examined in G.~Hiller,
  Phys.\ Rev.\ D {\bf 66}, 071502 (2002); A.~K.~Giri and R.~Mohanta,
  Phys.\ Rev.\ D {\bf 68}, 014020 (2003); D.~Atwood and G.~Hiller,
  arXiv:hep-ph/0307251; V.~Barger, C.~W.~Chiang, P.~Langacker and
  H.~S.~Lee, Phys.\ Lett.\ B {\bf 580}, 186 (2004); N.~G.~Deshpande
  and D.~K.~Ghosh, arXiv:hep-ph/0311332;

\bibitem{Bpenguin} A.~Datta and D.~London, arXiv:hep-ph/0403165.

\bibitem{alphapaper} In the SM, the decays $B_{d,s}^0 \to K^0\kbar$
can be used to obtain the CKM phase $\alpha$, see A.~Datta and
D.~London, Phys.\ Lett.\ B {\bf 533}, 65 (2002).

\bibitem{BDK} M. Gronau and D. Wyler, \plb{265}{1991}{172}; D. Atwood,
I. Dunietz and A. Soni, \prl{78}{1997}{3257}. See also M. Gronau and
D. London, \plb{253}{1991}{483}; I. Dunietz, \plb{270}{1991}{75};
N. Sinha and R. Sinha, \prl{80}{1998}{3706}.

\bibitem{Bpipi} M. Gronau and D. London, \prl{65}{1990}{3381}.

\bibitem{Brhopi} A.E. Snyder and H.R. Quinn, \prd{48}{93}{2139};
H.R. Quinn and J.P. Silva, \prd{62}{2000}{054002}.

\bibitem{DILPSS} A. Datta, M. Imbeault, D. London, V. Pag\'e, N. Sinha
and R. Sinha, in preparation.

\end{thebibliography}
\end{document}